\documentclass[a4paper]{article}
\usepackage{graphicx}
\usepackage[compress]{cite}
\usepackage{braket}
\usepackage{bm}
\usepackage{comment}
\usepackage{here}
\usepackage{a4wide}
\usepackage{authblk}
\usepackage{amsmath,amssymb}

\usepackage{float}

\begin{document}
\def\Journal#1#2#3#4{{#1} {\bf #2}, #3 (#4)}
\def\AHEP{Advances in High Energy Physics.} 	
\def\ARNPS{Annu. Rev. Nucl. Part. Sci.} 
\def\AandA{Astron. Astrophys.} 
\def\ANP{Ann. Phys.}
\def\APJ{Astrophys. J.}
\def\APJS{Astrophys. J. Suppl}
\def\COMR{Comptes Rendues}
\def\CQG{Class. Quantum Grav.}
\def\CPC{Chin. Phys. C}
\def\EPJC{Eur. Phys. J. C}
\def\EPL{EPL}
\def\FP{Fortsch. Phys.}
\def\IJMPA{Int. J. Mod. Phys. A}
\def\IJMPE{Int. J. Mod. Phys. E}
\def\JCAP{J. Cosmol. Astropart. Phys.}
\def\JHEP{J. High Energy Phys.}
\def\JETPL{JETP. Lett.}
\def\JETPUSSR{JETP (USSR)}
\def\JPG{J. Phys. G} 
\def\JPCS{J. Phys. Conf. Ser.} 
\def\JPGNP{J. Phys. G: Nucl. Part. Phys.} 
\def\JMP{J. Mod. Phys.} 
\def\MPLA{Mod. Phys. Lett. A}
\def\NIMA{Nucl. Instrum. Meth. A.}
\def\NATU{Nature}
\def\NCA{Nuovo Cimento}
\def\NJP{New. J. Phys.}
\def\NPB{Nucl. Phys. B}
\def\NPBOLD{Nucl. Phys.}
\def\NPBSUPPL{Nucl. Phys. B. Proc. Suppl.}
\def\PL{Phys. Lett.}
\def\PLB{{Phys. Lett.} B}
\def\PMCA{PMC Phys. A}
\def\PREP{Phys. Rep.}
\def\PPNP{Prog. Part. Nucl. Phys.}
\def\PLBOLD{Phys. Lett.}
\def\PAN{Phys. Atom. Nucl.}
\def\PRL{Phys. Rev. Lett.}
\def\PRD{Phys. Rev. D}
\def\PRC{Phys. Rev. C}
\def\PR{Phys. Rev.}
\def\PTP{Prog. Theor. Phys.}
\def\PTEP{Prog. Theor. Exp. Phys.}
\def\RMF{Revista Mexicana de Fisica.}
\def\RMP{Rev. Mod. Phys.}
\def\RPP{Rep. Prog. Phys.}
\def\SJNP{Sov. J. Nucl. Phys.}
\def\SPJETP{Sov. Phys. JETP.}
\def\SCIENCE{Science}
\def\TNYAS{Trans. New York Acad. Sci.}
\def\ZETP{Zh. Eksp. Teor. Piz.}
\def\ZFPH{Z. fur Physik}
\def\ZPC{Z. Phys. C}
\title{Modified TM$_2$ for Reproducing All Best-Fit Values of Neutrino Mixing Angles}
\author[1]{Michael Fodroci\footnote{5MTAD005@tokai.ac.jp}}
\author[2]{Teruyuki Kitabayashi \footnote{teruyuki@tokai.ac.jp}}
\affil[1]{Graduate School of Science and Technology, Tokai University, 4-1-1 Kitakaname, Hiratsuka, Kanagawa 259-1292, Japan}
\affil[2]{Department of Physics, School of Science, Tokai University, 4-1-1 Kitakaname, Hiratsuka, Kanagawa 259-1292, Japan}
\date{}
\maketitle
\begin{abstract}

As measurements of neutrino mixing angles continue to become more precise, it is increasingly likely that in the very near future a realistic neutrino mixing model will be required to precisely reproduce their best-fit values. In this study, a modified TM$_2$ mixing model which reproduces the best-fit values of all three neutrino mixing angles is proposed. The model reproduces the correct mixing angles within 1$\sigma$ of the current best-fit values and is robust against any future changes of the best-fit values.
\end{abstract}
\section{Introduction}
One of the long-standing unsolved problems in neutrino physics is the construction of a mixing model that correctly reproduces the three neutrino mixing angles obtained in neutrino oscillation experiments. Numerous models of the neutrino mixings and the flavor masses have been proposed previously, such as tri-bi maximal mixing (TBM) \cite{Harrison2002PLB,Xing2002PLB,Harrison2002PLB2,Kitabayashi2007PRD}, texture zeros \cite{Berger2001PRD,Frampton2002PLB,Xing2002PLB530,Xing2002PLB539,Kageyama2002PLB,Xing2004PRD,Grimus2004EPJC,Low2004PRD,Low2005PRD,Grimus2005JPG,Dev2007PRD,Xing2009PLB,Fritzsch2011JHEP,Kumar2011PRD,Dev2011PLB,Araki2012JHEP,Ludle2012NPB,Lashin2012PRD,Deepthi2012EPJC,Meloni2013NPB,Meloni2014PRD,Dev2014PRD,Felipe2014NPB,Ludl2014JHEP,Cebola2015PRD,Gautam2015PRD,Dev2015EPJC,Kitabayashi2016PRD1,Zhou2016CPC,Singh2016PTEP,Bora2017PRD,Barreiros2018PRD,Kitabayashi2018PRD,Barreiros2019JHEP,Capozzi2020PRD,Singh2020EPL,Barreiros2020,Kitabayashi2020PRD,Kitabayashi2017IJMPA,Kitabayashi2017IJMPA2,Kitabayashi2019IJMPA}, $\mu-\tau$ symmetric texture \cite{Fukuyama1997,Lam2001PLB,Ma2001PRL,Balaji2001PLB,Koide2002PRD,Kitabayashi2003PRD,Koide2004PRD,Aizawa2004PRD,Ghosal2004MPLA,Mohapatra2005PRD,Koide2005PLB,Kitabayashi2005PLB,Haba206PRD,Xing2006PLB,Ahn2006PRD,Joshipura2008EPJC,Gomez-Izquierdo2010PRD,He2001PRD,He2012PRD,Gomez-Izquierdo2017EPJC,Fukuyama2017PTEP,Kitabayashi2016IJMPA,Kitabayashi2016PRD,Bao2021arXiv,Garces2018JHEP,JuanCarlos2019JHEP,JuanCarlos2008PRD,Ge2010JCAP,He2015PLB,JuanCarlos2017IJMPA,Spinrath2012NPB,JuanCarlos2023RMF,Hyodo2025MPLA}, as well as textures and mixings under discrete symmetries, e.g., $A_n$ and $S_n$\cite{Altarelli2010PMP}. 

TM$_2$ mixing is one of the simplest neutrino mixing models\cite{Bjorken2006PRD,He2007PLB,Albright2009EPJC,Albright2010EPJC,He2011PRD,Kumar2010PRD,Gautam2016PRD,Chen2023IJMPA,Hyodo2024MPLA} which has the appealing feature that the mass matrix derived from the mixing matrix has a magic texture \cite{Harrison2004PLB,Lam2006PLB,Hyodo2020IJMPA,Yang2022PTEP,Channey2019JPGNP,Verma2020JPGNP,Hyodo2021PTEP, Verma2022arXiv}. However, some of the neutrino mixing angles predicted by the TM$_2$ model are only marginally within the $3\sigma$ region of the observed values. For this reason, there have been numerous attempts to modify the TM$_2$ mixing model. Each method of modification has its own unique advantages. For example, the modification in Ref.\cite{Hyodo2024MPLA} is devised to reproduce two of the three best-fit values,  those of the solar and reactor neutrino mixing angles, simultaneously. Because measurements of the neutrino mixing angles are becoming extremely precise, reproducibility of the best-fit values of neutrino mixing angles may be more important for a realistic neutrino mixing model in the very near future.

In this study, we show that not only two, but rather all three best-fit values of the neutrino mixing angles can be simultaneously reproduced via a modified TM$_2$ mixing model - obtained via the same method as in Ref.\cite{Hyodo2024MPLA}. Additionally, we show this reproducibility will persist under future changes to the best-fit values.

This paper is arranged as follows. In section \ref{section:Basics of neutrino mixings and masses} the basics of neutrino mixing matrices and mass matrices are given alongside a review of the current best-fit values of the various mixing parameters. In sections \ref{subsection:problem_within_TM_2_mixing} through \ref{subsubsection:Predictions of Modified TM2} we present our modification to the TM$_2$ mixing model and demonstrate its ability to precisely reproduce all three mixing angles as well as demonstrate how the model accomplishes this. In section \ref{subsubsection:Majorana CP phases} we discuss the Majorana CP phases and the ranges of the effective Majorana mass predicted by our model. In section \ref{subsubsection_magic_texture} we discuss the way in which in the deviation of our model from the standard TM$_2$ mixing matrix breaks the magic texture. Lastly, in section 4, we summarize our findings. 

\section{Basics of neutrino mixings and masses\label{section:Basics of neutrino mixings and masses}}
The neutrino sector can be parametrized by three mixing angles $\{ \theta_{12}, \theta_{23}, \theta_{13} \}$, one Dirac CP phase $\delta$, two Majorana CP phases $\{ \alpha, \beta \}$, and three neutrino mass eigenstates $\{m_1, m_2, m_3\}$. 

Theoretically, the neutrino mixings are parametrized by the Pontecorvo-Maki-Nakagawa-Sakata (PMNS) mixing matrix \cite{Pontecorvo1957,Pontecorvo1958,Maki1962PTP,PDG} 
\begin{align}
U  
&=\left(
\begin{matrix}
U_{e1} & U_{e2} & U_{e3}  \\
U_{\mu 1} & U_{\mu 2} & U_{\mu 3}  \\
U_{\tau 1} & U_{\tau 2} & U_{\tau 3}  \\
\end{matrix} 
\right)
\nonumber \\
&=
\left ( 
\begin{array}{ccc}
c_{12}c_{13} & s_{12}c_{13} & s_{13} e^{-i\delta} \\
- s_{12}c_{23} - c_{12}s_{23}s_{13} e^{i\delta} & c_{12}c_{23} - s_{12}s_{23}s_{13}e^{i\delta} & s_{23}c_{13} \\
s_{12}s_{23} - c_{12}c_{23}s_{13}e^{i\delta} & - c_{12}s_{23} - s_{12}c_{23}s_{13}e^{i\delta} & c_{23}c_{13} \\
\end{array}
\right).
\end{align}
Above we have used the conventional abbreviations $c_{ij}=\cos\theta_{ij}$, and $s_{ij}=\sin\theta_{ij}$, where ($i,j=\:1\:,2\:,3\:$). The sines and cosines of the three mixing angles can be obtained as follows:
\begin{align}
s_{12}^2 = \frac{|U_{e2}|^2}{1-|U_{e3}|^2},  \
s_{23}^2 = \frac{|U_{\mu 3}|^2}{1-|U_{e3}|^2},   \
s_{13}^2 = |U_{e3}|^2.  
\label{Eq:s12s_s23s_s13s} 
\end{align}
The Jarlskog invariant, $J_{\rm CP}$, \cite{Jarlskog1985PRL} is related to the Dirac CP phase, $\delta$, and, in the standard parameterization, may be expressed as follows:
\begin{equation}
\label{Eq:JCP}
J_{\rm CP} = {\rm Im} (U_{e1} U_{e2}^* U_{\mu 1}^* U_{\mu 2}) =  s_{12}s_{23}s_{13}c_{12}c_{23}c_{13}^2 \sin\delta.  
\end{equation}

The flavor neutrino mass matrix 
\begin{equation}
M_\nu
=\left( \begin{array}{ccc}
M_{ee} & M_{e\mu} & M_{e\tau} \\
M_{\mu e} & M_{\mu\mu} & M_{\mu\tau} \\
M_{\tau e} & M_{\tau\mu} & M_{\tau\tau}\\ 
\end{array} \right) 
\end{equation}
is obtained via 
\begin{equation}
M_\nu = U^* M_{\rm diag} U^\dag,
\end{equation}

%
\noindent where  
\begin{equation}
M_{\rm diag} = {\rm diag}(m_1, m_2e^{2i\alpha}, m_3e^{2i\beta})= {\rm diag}(m_1, \tilde{m}_2, \tilde{m}_3)
\end{equation}
is the diagonal neutrino mass matrix.

A global analysis of the current data obtained from neutrino oscillation experiments yields the following best-fit values in the normal mass ordering (NO) scenario, $m_1 < m_2 < m_3$, of the mixing angles and the squared mass differences $\Delta m_{ij}^2 = m^2_i - m^2_j$\cite{Esteban2024JHEP}:
\begin{align} 
s_{12}^2 &= 0.308^{+0.012}_{-0.011} \ (0.275 \rightarrow 0.345), \quad
s_{23}^2 = 0.470^{+0.017}_{-0.013} \ (0.435 \rightarrow 0.585 ), \nonumber\\
s_{13}^2 &= 0.02215^{+0.00056}_{-0.00058} \ (0.02030 \rightarrow 0.02388),\quad
\delta/^\circ = 212^{+26}_{-41} \ (124 \rightarrow 364),\nonumber \\
\frac{\Delta m_{21}^2}{10^{-5} {\rm ~ eV^2}} &= 7.49^{+0.19}_{-0.19} \ (6.92 \rightarrow 8.05), \quad
\frac{\Delta m_{31}^2}{10^{-3} {\rm ~ eV^2}} = 2.513^{+0.021}_{-0.019} \ (2.451 \rightarrow 2.578),  
\label{Eq:NuFit_NO}
\end{align}
where the $\pm$ represents the $1 \sigma$ region and the parentheses denote the $3 \sigma$ region.  For the inverted mass ordering (IO) scenario, $m_3 < m_1 \simeq m_2$, 
\begin{align}
s_{12}^2 &= 0.308^{+0.012}_{-0.011} \ (0.275 \rightarrow 0.345), \quad
s_{23}^2 = 0.550^{+0.012}_{-0.015} \ (0.440 \rightarrow 0.584), \nonumber\\
s_{13}^2 &= 0.02231^{+0.00056}_{-0.00056} \ (0.02060 \rightarrow 0.02409),\quad
\delta^\circ = 274^{+22}_{-25} \ (201 \rightarrow 335),\nonumber \\
\frac{\Delta m_{21}^2}{10^{-5} {\rm ~ eV^2}} &= 7.49^{+0.19}_{-0.19} \ (6.92 \rightarrow 8.05),\quad
\frac{\Delta m_{32}^2}{10^{-3} {\rm ~ eV^2}} = -2.484^{+0.020}_{-0.020} \ (-2.547 \rightarrow -2.421).
\label{Eq:NuFit_IO}
\end{align}
%
\section{Modified TM$_2$ mixing\label{section:Modified_TM2_mixing}}
\subsection{Problem within TM$_2$ mixing\label{subsection:problem_within_TM_2_mixing}}
The TM$_2$ mixing matrix is parametrized as 
\begin{equation}
U_{\rm TM_2}=\left(
\begin{matrix}
\sqrt{\frac{2}{3}}\cos\theta &  \sqrt{\frac{1}{3}} & \sqrt{\frac{2}{3}}\sin\theta  \\
-\frac{\cos\theta}{\sqrt{6}} + \frac{e^{-i\phi}\sin\theta}{\sqrt{2}}  &\sqrt{\frac{1}{3}}  & -\frac{\sin\theta}{\sqrt{6}} - \frac{e^{-i\phi}\cos\theta}{\sqrt{2}} \\
-\frac{\cos\theta}{\sqrt{6}} - \frac{e^{-i\phi}\sin\theta}{\sqrt{2}}  & \sqrt{\frac{1}{3}}  & -\frac{\sin\theta}{\sqrt{6}}  + \frac{e^{-i\phi}\cos\theta}{\sqrt{2}} \\
\end{matrix}
\right)
\label{Eq:UTM2}
\end{equation}
where $\theta$ and $\phi$ are real parameters. From Eq.(\ref{Eq:s12s_s23s_s13s}), we obtain the following stringent constraint between $s_{12}^2$ and $s_{13}^2$
\begin{align}
s^2_{12} &= \frac{1}{3(1-s^2_{13})}.
\end{align}
Since global analysis of neutrino oscillation data allows only a narrow $3\sigma$ region of $s_{13}^2$, predicted values via the TM$_2$ of $s_{12}^2$ should also reside in the same narrow region, i.e., $0.340 \le s^2_{12} \le 0.341$ for NO and $0.340 \le s^2_{12} \le 0.342$ for IO. Yet, the predicted value of $s^2_{12}$ almost approaches the upper limit of the $3\sigma$ allowed region. Hence, the TM$_2$ mixing model may be excluded from neutrino oscillation experiments in the future and modification of TM$_2$ is necessary.

\subsection{Modification\label{subsection:modfication}}
As was addressed in the introduction, the measurements of the neutrino mixing angles are becoming extremely precise. Therefore, the reproducibility of the best-fit values of neutrino mixing angles may become the first requirement for a neutrino mixing model in the very near future. In this context, we construct a modified TM$_2$ mixing model which can yield all three best-fit values of the mixing angles.


In accordance with the strategy of Ref.\cite{Hyodo2024MPLA}, we propose the following procedure for modifying the TM$_2$ mixing matrix.
\begin{enumerate}
\item First, because the original TM$_2$ mixing matrix is constructed from the TBM mixing matrix \cite{Harrison2002PLB,Xing2002PLB}, we start by modifying the TBM mixing matrix to improve the reproducibility of the solar mixing angle in a straightforward way.  
\item Next, we modify again the already modified TBM mixing matrix to enable the reproduction of the best-fit values of the reactor mixing angle. 
\end{enumerate}

As per the first step outlined above, we modify the TBM mixing matrix. The original TBM mixing is given to be:
\begin{eqnarray}
U_{\rm TBM}=\left(
\begin{array}{ccc}
\sqrt{\frac{2}{3}} & \sqrt{\frac{1}{3}}  & 0 \\
-\sqrt{\frac{1}{6}} &  \sqrt{\frac{1}{3}}  &-\sqrt{\frac{1}{2}}  \\
-\sqrt{\frac{1}{6}} &  \sqrt{\frac{1}{3}}  &\sqrt{\frac{1}{2}}
\end{array}
\right).
\label{Eq:UTBM}
\end{eqnarray}
Which yields the magnitude of $s_{12}^2$ 
\begin{eqnarray}
(s_{12}^2)_{\rm TBM}= |U_{e2}|^2 = \frac{1}{3}=0.3333,
\label{Eq:TBM_Predicted}
\end{eqnarray}
which does not match the best-fit values. Thus, we modify the TBM matrix to achieve the reproduction of the best-fit values for $s_{12}^2$. From Eq.(\ref{Eq:TBM_Predicted}), we see that $|U_{e2}|^2$ is directly related to values of $s_{12}^2$, and so we introduce the following modification to $U_{e2}$:
\begin{eqnarray}
\sqrt{\frac{1}{3}} \rightarrow \sqrt{\frac{1}{3}+\epsilon},
\end{eqnarray}
where $\epsilon$ denotes a real parameter. The best-fit value $s_{12}^2 = 0.303$ is reproduced for $\epsilon=-0.0303$ . This gives us the modified TBM  mixing matrix,
\begin{eqnarray}
U_{\rm MTBM}^{0}=\left(
\begin{array}{ccc}
\sqrt{\frac{2}{3}} & \sqrt{\frac{1}{3} + \epsilon}  & 0 \\
-\sqrt{\frac{1}{6}} &  \sqrt{\frac{1}{3}}  &-\sqrt{\frac{1}{2}}  \\
-\sqrt{\frac{1}{6}} &  \sqrt{\frac{1}{3}}  &\sqrt{\frac{1}{2}}
\end{array}
\right).
\label{Eq:UMTBM_non}
\end{eqnarray}
While it is now possible to derive the best-fit value of $s_{12}^2$ from $U_{\rm MTBM}^{0}$, this matrix is not a unitary. In order to guarantee the unitarity condition, we modify Eq. (\ref{Eq:UMTBM_non}) like so 
\begin{eqnarray}
U_{\rm MTBM}=\left(
\begin{array}{ccc}
\sqrt{\frac{2}{3} + x_1} & \sqrt{\frac{1}{3} + \epsilon}  & 0 \\
-\sqrt{\frac{1}{6}+ x_2} &  \sqrt{\frac{1}{3}+ x_3}  &-\sqrt{\frac{1}{2}+ x_4}  \\
-\sqrt{\frac{1}{6}+ x_5} &  \sqrt{\frac{1}{3}+ x_6}  &\sqrt{\frac{1}{2}+ x_7}
\end{array},
\right)
\label{Eq:UMTBM_unitarity}
\end{eqnarray}
where $x_i$ ($i=1,2,\cdots,7$) are real parameters. The unitary condition  $U_{\rm MTBM}^\dag U_{\rm MTBM}=1$ is equivalent to the following system of simultaneous equations 
\begin{eqnarray}
&& 1+x_1+\epsilon = 1, \quad 1+x_2+x_3+x_4=1,  \quad 1+x_5+x_6+x_7=1, \nonumber \\
&& -\sqrt{\left(\frac{2}{3}+x_1\right) \left(\frac{1}{6}+x_2\right)}+\sqrt{\left(\frac{1}{3}+x_3\right) \left(\frac{1}{3}+\epsilon\right)}=0,  \nonumber \\
&& -\sqrt{\left(\frac{2}{3}+x_1\right)\left(\frac{1}{6}+x_5\right)}+\sqrt{\left(\frac{1}{3}+x_6\right)\left(\frac{1}{3}+\epsilon\right)}=0,\nonumber \\
&& \sqrt{\left(\frac{1}{6}+x_2\right)\left(\frac{1}{6}+x_5\right)}+\sqrt{\left(\frac{1}{3}+x_3\right)\left(\frac{1}{3}+x_6\right)}-\sqrt{\left(\frac{1}{2}+x_4\right)\left(\frac{1}{2}+x_7\right)}=0. 
\end{eqnarray}
Since the number of independent equations is six but there are seven variables, the simultaneous equations cannot be solved in the current form. We use the fact that
\begin{eqnarray}
(s_{23}^2)_{\rm TBM}=|U_{\mu3}|^2=\frac{1}{2}
\end{eqnarray}
is consistent with the observation to set
\begin{eqnarray}
x_4=0,\:\: x_7=0,
\end{eqnarray}
thus reducing the number of unkowns and allowing us to solve the simultaneous equations. In this case we obtain
\begin{eqnarray}
(x_1,x_2,x_3,x_4,x_5,x_6,x_7) =\left (-\epsilon,\frac{\epsilon}{2},-\frac{\epsilon}{2},0,\frac{\epsilon}{2},-\frac{\epsilon}{2},0\right),
\label{Eq:solution}
\end{eqnarray}
and Eq.(\ref{Eq:UMTBM_non}) becomes
\begin{eqnarray}
U_{\rm MTBM}=\left(
\begin{array}{ccc}
\sqrt{\frac{2}{3} - \epsilon} & \sqrt{\frac{1}{3} + \epsilon}  & 0 \\
-\sqrt{\frac{1}{6}+ \frac{\epsilon}{2}} &  \sqrt{\frac{1}{3}-\frac{\epsilon}{2}}  &-\sqrt{\frac{1}{2}}  \\
-\sqrt{\frac{1}{6}+ \frac{\epsilon}{2}} &  \sqrt{\frac{1}{3}-\frac{\epsilon}{2}}  &\sqrt{\frac{1}{2}}
\end{array}
\right).
\label{Eq:UMTBM}
\end{eqnarray}
The range of $\epsilon$ should be
\begin{eqnarray}
-\frac{1}{3} < \epsilon < \frac{2}{3}
\label{Eq:a_region_1}
\end{eqnarray}
because the contents of the square roots cannot be negative. Moreover, we naturally expect that, as a perturbation, $|\epsilon| \ll 1$ . We would like to note that the modified TBM mixing in Eq.(\ref{Eq:UMTBM}) can be regarded as an example of a minimally modified TBM mixing matrix. 

The TBM mixing structure already takes care of the observed maximal $\theta_{23}$ and it is necessary to maintain this situation as much as possible if we modify the TBM. In this study, the maximal $\theta_{23}$ is guaranteed by $x_4=0$ under TBM mixing. Moreover, the TBM mixing structure already takes care of the modestly large $\theta_{12}$. Furthermore, although the predicted $\theta_{12}$ does not match the best-fit value, the situation where $\theta_{12}$ is modestly large should be maintained. With this in mind, we take the straightforward modification of $\theta_{12}$ where $|U_{e2}|^2= \frac{1}{3}+\epsilon$ which is based on the relation $(s_{12}^2)_{\rm TBM}= |U_{e2}|^2 = \frac{1}{3}$. After these two ingredients are taken into account, we obtain the final texture of the modified TBM mixing which was obtained in Eq.(\ref{Eq:UMTBM}) via imposing the unitary condition.

According to the second step outlined at the beginning of the section, we modify again the already modified TBM mixing to enable the reproduction of the best-fit values for the reactor mixing angle (the magnitude of $s_{13}^2$ obtained from the modified TBM, $(s_{13}^2)_{\rm TBM}=|U_{e3}|^2 = 0$, being inconsistent with observations). The original TM$_2$ mixing matrix is established as that for which the second column of the TBM mixing matrix remains unaltered, while the first and third columns are modified. Similarly, we contemplate a mixing matrix that retains the second column of $U_{\rm MTBM}$ unchanged as follows

%
\begin{equation}
\tilde{U}_{\rm TM_2} =\left(
\begin{matrix}
\sqrt{\frac{2}{3}-\epsilon}\cos\theta &  \sqrt{\frac{1}{3}+\epsilon} & \sqrt{\frac{2}{3}-\epsilon}\sin\theta  \\
-\sqrt{\frac{1}{6}+\frac{\epsilon}{2}}\cos\theta + \frac{e^{-i\phi}\sin\theta}{\sqrt{2}}  &\sqrt{\frac{1}{3}-\frac{\epsilon}{2}}  & -\sqrt{\frac{1}{6}+\frac{\epsilon}{2}}\sin\theta - \frac{e^{-i\phi}\cos\theta}{\sqrt{2}} \\
-\sqrt{\frac{1}{6}+\frac{\epsilon}{2}}\cos\theta - \frac{e^{-i\phi}\sin\theta}{\sqrt{2}}  & \sqrt{\frac{1}{3}-\frac{\epsilon}{2}}  & -\sqrt{\frac{1}{6}+\frac{\epsilon}{2}}\sin\theta + \frac{e^{-i\phi}\cos\theta}{\sqrt{2}} \\
\end{matrix}
\right).
\label{Eq:UMTM2}
\end{equation}
This is the modified TM$_2$ mixing matrix whose investigation is the focus of this paper. We note that, as expected, the modified TM$_2$ mixing matrix is unitary, and that the original TM$_2$ mixing is reproduced if $\epsilon=0$, and furthermore that the original TBM mixing matrix is reproduced if $\epsilon = \phi= \theta = 0$.

The mixing angles and Dirac CP phase predicted by $\tilde{U}_{\rm TM_2}$ are 
\begin{align}
s^2_{12} &=  \frac{1+3\epsilon}{3-(2-3\epsilon)\sin^2\theta},  \label{Eq:s12sMTM2} \\
s^2_{23} & = \frac{1}{2}\left( 1+\frac{\sqrt{3(1+3\epsilon)}\sin2\theta \cos\phi}{3-(2-3\epsilon)\sin^2\theta} \right),  \label{Eq:s23sMTM2}\\
s^2_{13} &= \left(\frac{2}{3}-\epsilon\right)\sin^2\theta,  \label{Eq:s13sMTM2}\\
\tan \delta &= \frac{4+3\epsilon +(2-3\epsilon) \cos2\theta}{2-3\epsilon + (4+3\epsilon)\cos2\theta}\tan\phi.  \label{Eq:deltaMTM2}
\end{align}
%


\subsection{Properties of the modified TM$_2$\label{Subsection:Properties_of_the_modified_TM2}}
\subsubsection{Reproduction of the current best-fit values of mixing angles\label{subsubsection:Reproduction of the current best-fit values of mixing angles}}
The modified TM$_2$ mixing has three parameters $\theta \simeq \theta_{13}$, $\phi$, and $\epsilon$, and thus we can naturally expect that best-fit values of the three mixing angles can be fit by these three model parameters.  Indeed, a benchmark point
\begin{equation}
(\theta, \phi, \epsilon) = (167.76^\circ, 287.62^\circ, -0.03216)
\label{NO_theta_phi_epsilon}
\end{equation}
yields
\begin{equation}
(s^2_{12}, s^2_{23}, s^2_{13}, \delta) = (0.308, 0.470,0.02215,287.6^\circ)
\label{NO_Predicted_Angles}
\end{equation}
which perfectly matches the best-fit values for all of three mixing angles and is consistent with the $3\sigma$ allowed region of the Dirac CP phase for NO. Another benchmark point 
\begin{equation}
(\theta, \phi, \epsilon) = (169.71^\circ, 239.50^\circ, -0.0322)
\label{IO_theta_phi_epsilon}
\end{equation}
yields
\begin{equation}
(s^2_{12}, s^2_{23}, s^2_{13}, \delta) = (0.308, 0.550,0.02231,239.5^\circ)
\label{IO_Predicted_Angles}
\end{equation}
which is again perfectly matches the best-fit values for all of three mixing angles and is consistent with the $3\sigma$ allowed region of the Dirac CP phase for IO. 


\subsubsection{Predictions of Modified TM$_2$ Model Around NuFIT 3$\sigma$ Allowed Regions\label{subsubsection:Predictions of Modified TM2}}

While it is clear that the parameters of the modified TM$_2$ mixing model may be tuned to match best-fit data, one may wonder how exactly it manages to reproduce the observed mixing angles. It is also interesting to see for what region of parameter space the values predicted by the modified TM$_2$ model remain close to the current best-fit values. In order to investigate this we present six different plots in Figures 1 and \ref{IO_Triple} wherein we have held fixed one of the three mixing angles while varying $\theta$/$\epsilon$/$\phi$ in both the NO and IO cases. The predictions of our model under various ranges of parameters are then overlaid with the 1/2/3$\sigma$ regions given by NuFit\cite{Esteban2024JHEP}

In the left hand figures we set $\sin^2\theta_{12}$ equal to its best-fit value as per Eqs.(\ref{Eq:NuFit_NO}) and (\ref{Eq:NuFit_IO}) as well as set $\phi$ to the values in Eqs.(\ref{NO_theta_phi_epsilon}) and (\ref{IO_theta_phi_epsilon}). We then vary $\epsilon$ between $-\frac{1}{30}$ and $\frac{1}{30}$ and use Eq.(\ref{Eq:s12sMTM2}) to compute $\theta$ - only keeping solutions that fall in the range $\frac{3}{4}\pi\leq\theta\leq\pi$. This is then repeated for several reasonable values of $\phi$ around the intial value. For the middle figures, $\sin^2\theta_{13}$ is fixed at the best-fit value instead and the process repeated. 

For the right most figures corresponding to a fixed $\sin^2\theta_{23}$ the situation is not as simple. Inspection of Eq.(\ref{Eq:s23sMTM2}) shows this same process cannot be repeated as easily due to the more complicated form of this equation. Therefore, in order to proceed with the analysis we take Eq.(\ref{Eq:s23sMTM2}) and rewrite it as follows.

\begin{align}
s^2_{23}  = \frac{1}{2}\left( 1+\frac{\sqrt{3}\sqrt{1+3\epsilon}\sin2\theta \cos\phi}{3-(2-3\epsilon)\sin^2\theta} \right)
\end{align}

\noindent Using the fact that $3\epsilon$ is a small parameter we replace the root containing $\epsilon$ with its Taylor series expansion. 

\begin{align}
s^2_{23}  = \frac{1}{2}\left( 1+\frac{\sqrt{3}(1+\frac{3\epsilon}{2}-\frac{9\epsilon^2}{8}+\:\:...\:\:)\sin2\theta \cos\phi}{3-(2-3\epsilon)\sin^2\theta} \right)
\end{align}

\noindent Truncating at the third order term leaves us with a much more manageable expression which is quadratic in $\epsilon$. 

\begin{align}
s^2_{23} & = \frac{1}{2}\left( 1+\frac{\sqrt{3}(1+\frac{3\epsilon}{2}-\frac{9\epsilon^2}{8})\sin2\theta \cos\phi}{3-(2-3\epsilon)\sin^2\theta} \right)
\end{align}

On the relevant range of $\epsilon$ the error at the third order can be shown to be less than 1\%. Next, using this new experssion, we again set $\phi$ to the values from Eqs.(\ref{NO_theta_phi_epsilon})/(\ref{IO_theta_phi_epsilon}) and then vary $\theta$ between $\frac{3}{4}\pi$ and $\pi$. We solve the quadratic at each value of $\theta$ and keep only the solutions for which $\epsilon$ falls between $-\frac{1}{30}$ and  $\frac{1}{30}$. This is the reverse of the left and middle plots where we plugged in a value for $\epsilon$ to compute $\theta$. This is again repeated for several reasonable values of $\phi$ around the initial value.

\begin{figure}[H]
    \centering
    \includegraphics[width=1\linewidth]{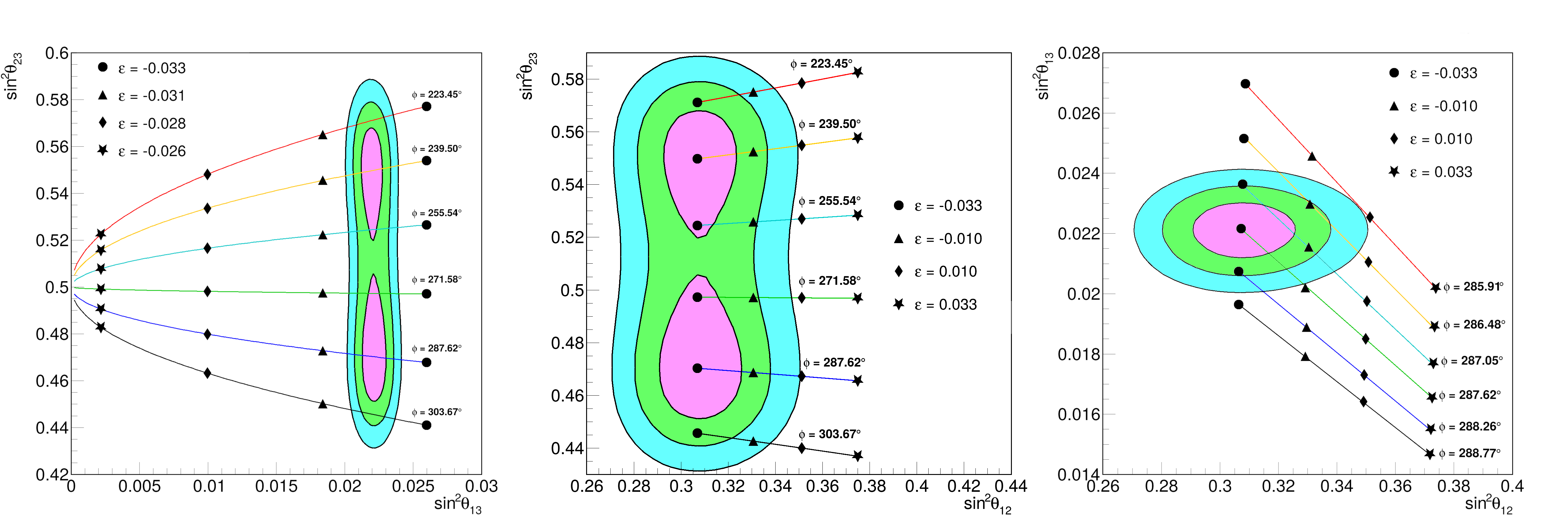}
    \caption{NO case predictions of the modified TM$_2$ model for fixed $\sin^2\theta_{12}$(left), fixed $\sin^2\theta_{13}$(middle), and fixed $\sin^2\theta_{23}$(right). Shaded regions correspond to NuFIT\cite{Esteban2024JHEP} $1\sigma$(magenta), $2\sigma$(green), and $3\sigma$(cyan) allowed regions.  Points of the same shape indicated equivalent values of $\epsilon$ while the various colored curves indicate level lines of $\phi$. Points are restricted to $-\frac{1}{30}\leq\epsilon\leq\frac{1}{30}$ and $\frac{3}{4}\pi\leq\theta\leq\pi$.}
   \label{NO_triple}
\end{figure}

\begin{figure}[H]
    \centering
    \includegraphics[width=1\linewidth]{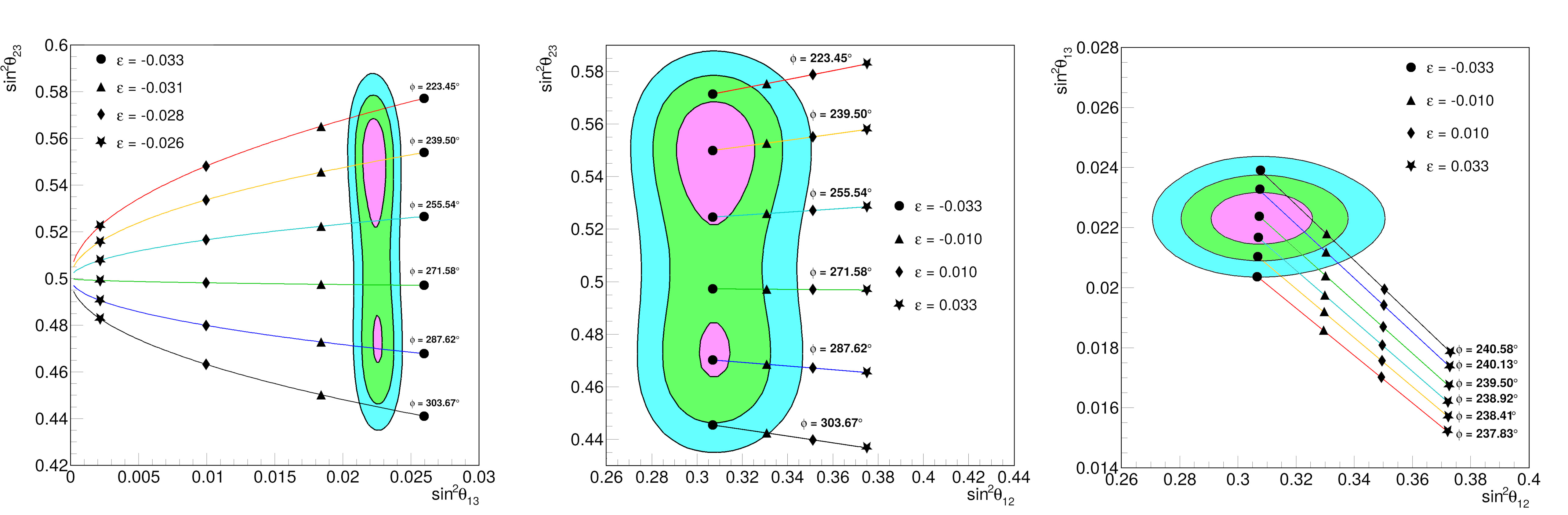}
    \caption{IO case predictions of the modified TM$_2$ model for fixed $\sin^2\theta_{12}$(left), fixed $\sin^2\theta_{13}$(middle), and fixed $\sin^2\theta_{23}$(right). Shaded regions correspond to NuFIT\cite{Esteban2024JHEP} $1\sigma$(magenta), $2\sigma$(green), and $3\sigma$(cyan) allowed regions. Points of the same shape indicated equivalent values of $\epsilon$ while the various colored curves indicate level lines of $\phi$. Points are restricted to $-\frac{1}{30}\leq\epsilon\leq\frac{1}{30}$ and $\frac{3}{4}\pi\leq\theta\leq\pi$.}
    \label{IO_Triple}
\end{figure}

It should be noted that for the IO and NO case in the left/middle plots (fixed $\sin^2\theta_{12}$/$\sin^2\theta_{13}$) that the same set of $\phi$ values was able to be used. This has partly to do with the very small differences between the NO best-fit values of $\sin^2\theta_{12}$/$\sin^2\theta_{13}$ and their IO best-fit values. Furthermore, those angles both lack $\phi$ dependence (see Eqs.(\ref{Eq:s12sMTM2}) and (\ref{Eq:s13sMTM2})). On the other hand, for $\sin^2\theta_{23}$, there is a large difference between the best-fit values in the NO and IO cases. This leads to the appropriate ranges of $\phi$ being very distinct between the rightmost NO and IO plots. Furthermore in those fixed $\sin^2\theta_{23}$ plots, the appropriate ranges of $\phi$ for NO and IO exist on either side of $\phi=\frac{3}{2}\pi$. This results in the relationship between $\phi$ and $\sin^2\theta_{13}$ flipping on either side of this value. This is why the smallest value of $\phi$ appears at the top of the NO plot, while appearing at the bottom of the IO plot. Accordingly, in the left/middle plots (fixed $\sin^2\theta_{12}$/$\sin^2\theta_{13}$) the sign of the slopes of the $\phi$ level lines on either side of $\phi=\frac{3}{2}\pi$ flip. For this to be clear it is also important to note that, for all values of the model parameter $\theta$ considered, $\sin(2\theta)<0$.

From these plots we may conlcude that the modified TM$_2$ model can accomdate the results of future precision measurements by adopting suitable values of three parameters $\theta$, $\phi$, and $\epsilon$ if the newly measured values fall into the $3\sigma$ allowed regions of the current measurements. The most stringent constraint on $\epsilon$ comes from the leftmost plots(fixed $\sin^2\theta_{12}$), where, in order to fall within the $3\sigma$ allowed region it is required that $-0.033<\epsilon<-0.031$. On the other hand, the third set of plots(fixed $\sin^2\theta_{23}$), set the tightests restrictions on $\phi$ where in order to remain within the $3\sigma$ region, in the NO case, $\phi$ must fall approximately within $285.91^\circ<\phi<288.77^\circ$, while for IO this range is $237.83^\circ<\phi<240.58^\circ$. In other words, the quadrant of $\phi$ is determined by the mass ordering. Furthermore, in all cases, $\theta$ must fall within the range $168.45^\circ<\theta<183.35^\circ$. Due to the strict constraints on $\epsilon$ imposed by the fixed $\sin^2\theta_{12}$ case and on $\phi$ by the $\sin^2\theta_{23}$ case, it is clear that the region in which our model may reproduce points that match experimental data is confined to a small area about the best-fit values within the 1$\sigma$ region.  

\subsubsection{Majorana CP phases\label{subsubsection:Majorana CP phases}}
Theoretical prediction of the two Majorana CP phases is also an important issue in neutrino physics. These Majorana CP phases cannot be determined by experiments on neutrino oscillations, rather information about these phases may be obtained through neutrinoless double beta decay experiments. On the other hand, theoretical predictions of the Majorana CP phases are obtainable via the flavor neutrino mass matrix, which may be constructed from the neutrino mixing matrix.

The relationship between the effective mass of neutrino-less double beta decay and the Majorana phases for the currently observed mixing angles is well understood. Since the Modified TM2 model can be fit to the observed mixing angles, its predictions for the effective mass and Majorana phases should be identical to the known predictions. For the sake of completeness, we confirm this.

With respect to the modified TM$_2$ mixing in Eq.(\ref{Eq:UMTM2}), the Majorana CP phases and the effective Majorana neutrino mass for neutrinoless double beta decay are related through  
\begin{align}
m_{\beta\beta} &=\sum_i (\tilde{U}_{\rm TM_2})^2_{ei} m_i \nonumber \\
&=\frac{1}{9}\left\{ (2 - 3 \epsilon)^2 m_1^2  \cos^4\theta + (1 +  3  \epsilon)^2 m_2^2 + (2 -  3 \epsilon)^2 m_3^2 \sin^4\theta \right. \nonumber \\
& \quad \left. + 2 (2 + 3 \epsilon - 9 \epsilon^2) m_1 m_2 \cos^2\theta  \cos 2 \alpha + 2  (2 + 3 \epsilon - 9 \epsilon^2) m_2 m_3  \sin^2\theta \cos 2 (\alpha - \beta) \right. \nonumber \\
& \quad \left. + 2 (2 - 3 \epsilon)^2 m_1 m_3  \cos^2\theta \sin^2\theta \cos 2 \beta \right\}.
\end{align}
Because each of the $ei$ elements are independent of $\phi$ in $\tilde{U}_{\rm TM_2}$, $m_{\beta\beta}$ is also independent of $\phi$. 

If one of the neutrino masses vanishes, such as in the minimal seesaw model, only one physical Majorana CP phase survives. More concretely, we can take the diagonal neutrino mass matrix to be $M_{\rm diag}={\rm diag.}(0, m_2 e^{2i\alpha},m_3)$ for the NO case and $M_{\rm diag}={\rm diag.}(m_1, m_2 e^{2i\alpha}, 0)$ for IO. This way, the Majorana CP phase is estimated as
\begin{equation}
\cos 2\alpha= \frac{9 m_{\beta\beta}^2 -  (1 + 3  \epsilon)^2 m_2^2 - (2 - 3 \epsilon)^2 m_3^2  \sin^4\theta }{2  (2 + 3 \epsilon - 9 \epsilon^2) m_2 m_3 \sin^2\theta}
\label{Eq:cos_2_alpha_of_mbb_NO}
\end{equation}
for $m_1=0$ and  
\begin{equation}
\cos 2\alpha= \frac{9 m_{\beta\beta}^2 -  (1 + 3 \epsilon)^2 m_2^2 -  (2 - 3 \epsilon)^2 m_1^2 \cos^4\theta}{2   (2 + 3 \epsilon - 9 \epsilon^2) m_1 m_2\cos^2\theta}
\label{Eq:cos_2_alpha_of_mbb_IO}
\end{equation}
for $m_3=0$.

\begin{figure}[H]
    \centering
    \includegraphics[width=0.75\linewidth]{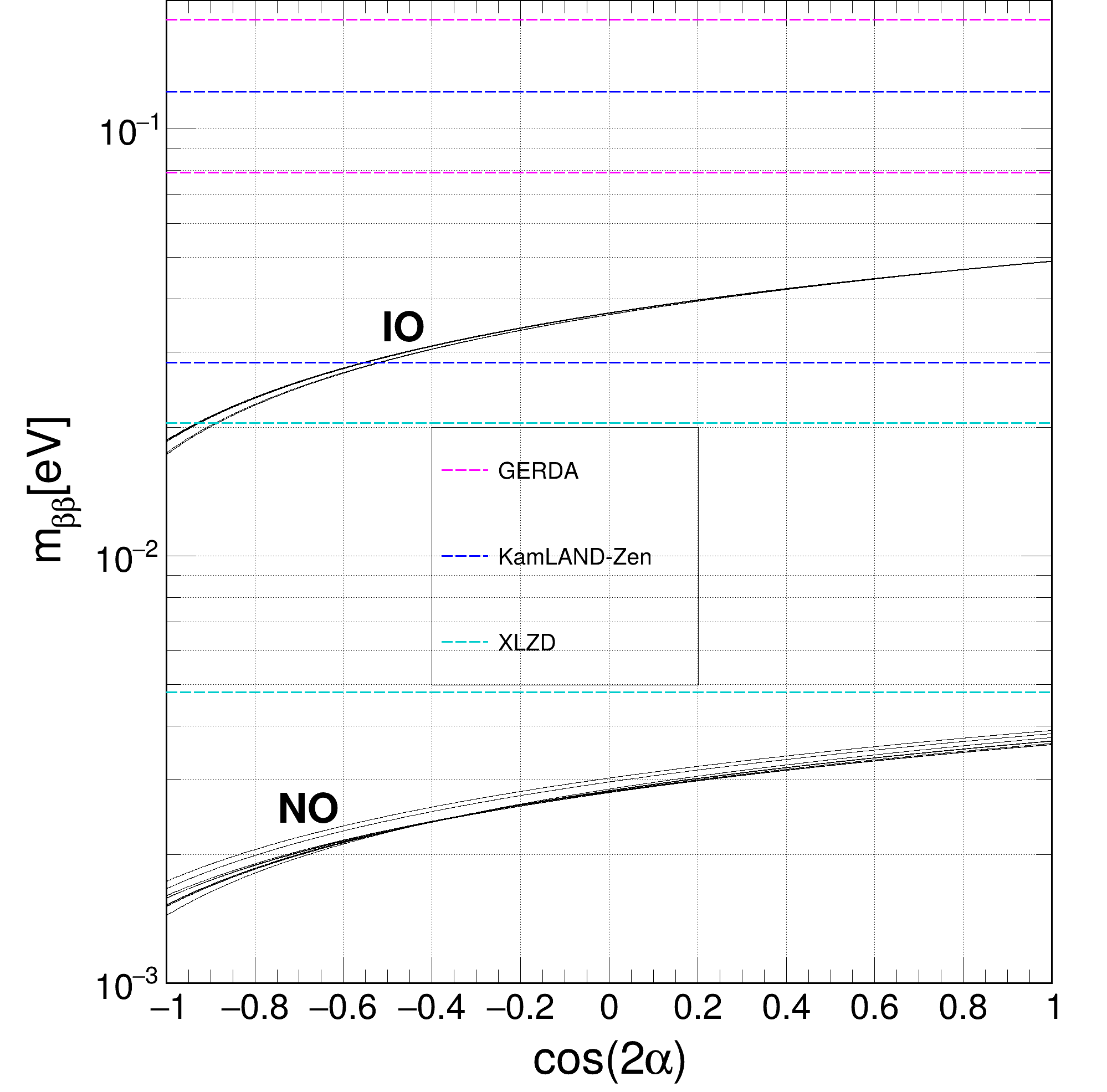}
    \caption{$m_{\beta\beta}$ vs $\cos(2\alpha)$ in the NO(black/bottom) and IO(black/top) scenarios. Parameters used to generate each curve correspond to the best fit point plus a sample from the 2 and 3$\sigma$ regions in each of the three fixed-angle scenrios. Also shown are the ranges on the upper limit of $m_{\beta\beta}$ from the full GERDA\cite{GERDA} dataset(dashed/magenta), the estimated range for the future XLZD\cite{XLZD} experiment(dashed/cyan), and from the full KamLAND-Zen\cite{KamLAND-Zen} dataset(dashed/dark blue) - all of which are at the 90\% C.L..}
    \label{fig:majorana_mass_vs_cos_2_alpha}
\end{figure}

In Figure \ref{fig:majorana_mass_vs_cos_2_alpha} we plot $m_{\beta\beta}$ versus $\cos2\alpha$ in the NO/IO case using the appropriate rearrangement of Eq.(\ref{Eq:cos_2_alpha_of_mbb_NO})/Eq.(\ref{Eq:cos_2_alpha_of_mbb_IO}). We have done this at 3 points for each of the graphs in Figures \label{NO_triple} and \label{IO_triple} - one point at the best-fit value as well as a point sampling the 2$\sigma$ and 3$\sigma$ allowed regions, all three of which are taken at distinct $\phi$ values. While, as noted, $m_{\beta\beta}$ is independent of $\phi$, this is not the case for the magic texture symmetry breaking parameter, $\Delta S$, which will be investigated using the same set of values in \ref{subsubsection_magic_texture}. In the NO case this corresponds to the following range of model parameters: 

\[ \mathrm{NO\:\:Fixed}\:\: \theta_{13}:\begin{cases} 
      169.56^\circ\leq\theta\leq169.75^\circ\\
      223.45^\circ\leq\phi\leq287.62^\circ  \\
-0.0333\leq\epsilon\leq-0.0079
   \end{cases},
\]
\[ \mathrm{NO\:\:Fixed}\:\: \theta_{12}:\begin{cases} 
      169.76^\circ\leq\theta\leq170.12^\circ\\
      223.45^\circ\leq\phi\leq287.62^\circ  \\
-0.0321\leq\epsilon\leq-0.0317
   \end{cases},
\]
\[ \mathrm{NO\:\:Fixed}\:\: \theta_{23}:\begin{cases} 
      169.56^\circ\leq\theta\leq169.75^\circ\\
      223.45^\circ\leq\phi\leq287.62^\circ  \\
-0.0325\leq\epsilon\leq-0.0331
   \end{cases}.
\]

\noindent While in the IO case this corresponds to:

\[ \mathrm{IO\:\:Fixed}\:\: \theta_{13}:\begin{cases} 
      169.61^\circ\leq\theta\leq169.72^\circ\\
      239.50^\circ\leq\phi\leq303.67^\circ  \\
-0.0332\leq\epsilon\leq-0.0199
   \end{cases},
\]
\[ \mathrm{IO\:\:Fixed}\:\: \theta_{12}:\begin{cases} 
      169.71^\circ\leq\theta\leq169.72^\circ\\
      239.50^\circ\leq\phi\leq303.67^\circ  \\
-0.0322\leq\epsilon\leq-0.0321
   \end{cases},
\]
\[ \mathrm{IO\:\:Fixed}\:\: \theta_{23}:\begin{cases} 
      169.50^\circ\leq\theta\leq170.17^\circ\\
      237.83^\circ\leq\phi\leq240.13^\circ  \\
-0.0331\leq\epsilon\leq-0.0330
   \end{cases}.
\]

Furthermore, in Figure \ref{fig:majorana_mass_vs_cos_2_alpha}, we have included several notable ranges of the upper limit on $m_{\beta\beta}$ from current and future experiments. In magenta we show the range taken from the final results of GERDA\cite{GERDA} at 90\% C.L.

\begin{equation}
m_{\beta\beta,\mathrm{\:GERDA}}<\:79\:[\mathrm{meV}]\:-\:180\:[\mathrm{meV}].
\end{equation}\newline

\noindent In dark blue we show the predicted 90\% C.L. range for the planned XLZD\cite{XLZD} experiment 

\begin{equation}
m_{\beta\beta,\mathrm{\:XLZD}}<\:4.8\:[\mathrm{meV}]\:-\:28.5\:[\mathrm{meV}].
\end{equation}\newline

\noindent Lastly, in cyan, we show the range of 90\% C.L. upper limits extrapolated from the full KamLAND-Zen\cite{KamLAND-Zen} dataset. We note here that the upper and lower values of this range come from different theoretical models employed in the analysis - full details are present in Ref.\cite{KamLAND-Zen}.

\begin{equation}
m_{\beta\beta,\mathrm{\:KamLAND-Zen}}<\:28.4\:[\mathrm{meV}]\:-\:122\:[\mathrm{meV}].
\end{equation}\newline

As was expected, Figure \ref{fig:majorana_mass_vs_cos_2_alpha} shows that the predictions of the modified TM$_2$ mixing model are in agreement with the notion that the next generation of $0\nu\beta\beta$ experiments should be able to exclude the IO region of allowed Majorana mass at high confidence, but not yet probe that of NO.

\subsubsection{Magic texture\label{subsubsection_magic_texture}}
As we noted in the introduction, the neutrino mass matrix derived using the TM$_2$ mixing matrix, 
\begin{equation}
M_{\rm TM_2} = U_{\rm TM_2}^* M_{\rm diag} U_{\rm TM_2}^\dag,
\end{equation}
acquires so-called magic texture, that is - the sum of the elements in any row or column of the neutrino mass matrix are identical \cite{Lam2006PLB, Harrison2004PLB}. A general matrix with magic texture has the form
\begin{equation}
M_{\rm magic}
=
\left( \begin{array}{ccc}
a & b & c \\
b & d & a+c-d \\
c & a+c-d & b-c+d\\ 
\end{array} \right).
\end{equation}
The identical value of the various sums is $a+b+c$. For $M_{\rm TM_2}$, the identical value is $\tilde{m}_2$. 

In the case of the modified TM$_2$ model, the neutrino mass matrix is given by 
\begin{equation}
\tilde{M}_{\rm TM_2} = \tilde{U}_{\rm TM_2}^* M_{\rm diag} \tilde{U}_{\rm TM_2}^\dag.
\end{equation}

Said modifications result in a breaking of this magic texture symmetry. This may be seen by examining the sums of the elements in the first, second, and third rows of $\tilde{M}_{\rm TM_2}$. These are given by
\begin{align}
S_1 &=  \frac{1}{3} \left(1  + \sqrt{4 + 6 (1 - 3 \epsilon) \epsilon} + 3 \epsilon \right) \tilde{m}_2 +\frac{1}{3} \left(2  - \sqrt{4 + 6 (1 - 3 \epsilon) \epsilon} - 3 \epsilon \right) (m_1 \cos^2\theta + \tilde{m}_3\sin^2\theta  ) ,
\nonumber \\
S_2 &= \frac{1}{6}  \left(4 + \sqrt{ 4 + 6 (1 - 3 \epsilon) \epsilon}- 6 \epsilon \right) \tilde{m}_2 + \frac{1}{6} \left(2  - \sqrt{ 4 + 6 (1 - 3 \epsilon) \epsilon}+ 6 \epsilon \right) (m_1 \cos^2\theta+\tilde{m}_3\sin^2\theta) 
\nonumber \\
& \quad +  \frac{\sqrt{3}}{12}   \left(\sqrt{4 - 6 \epsilon} -  2 \sqrt{1 + 3 \epsilon}\right) (m_1 - \tilde{m}_3)e^{i\phi} \sin{2 \theta},
\nonumber \\
\mathrm{and}
\nonumber \\
S_3 &= \frac{1}{6}  \left(4 + \sqrt{ 4 + 6 (1 - 3 \epsilon) \epsilon}- 6 \epsilon \right) \tilde{m}_2 + \frac{1}{6} \left(2  - \sqrt{ 4 + 6 (1 - 3 \epsilon) \epsilon}+ 6 \epsilon \right)(m_1 \cos^2\theta+\tilde{m}_3\sin^2\theta) 
\nonumber \\
& \quad -  \frac{\sqrt{3}}{12}   \left(\sqrt{4 - 6 \epsilon} -  2 \sqrt{1 + 3 \epsilon}\right) (m_1 - \tilde{m}_3)e^{i\phi} \sin{2 \theta},
\end{align}
respectively. For $\epsilon=0$ we recover $S_1=S_2=S_3=\tilde{m}_2$ as expected. 

Since our initial approach was to first consider small perturbations on the TBM mixing matrix characterized by the small parameter $\epsilon$, followed by aditional modifications characterized by $\theta$ and $\phi$, and since the original TBM matrix was in possesion of magic texture symmetry, we are interested in discovering cases in which the brokenness of this magic texture symmetry is minimized. For $\epsilon,\theta,\phi\neq0$ we may define, as a measure of the degree to which this magic texture symmetry is broken, the quantity 

\begin{equation}
    \Delta S \equiv 1 - \frac{|S_1|+|S_2|+|S_2|}{3|\tilde{m}_2|}.
\end{equation}

\noindent Furthermore, for notational simplicity, we define the terms $A$ through $E$ as

\begin{equation}
    A\equiv\frac{1}{3} \left(1  + \sqrt{4 + 6 (1 - 3 \epsilon) \epsilon} + 3 \epsilon \right),
\end{equation}

\begin{equation}
    B\equiv\frac{1}{3} \left(2  - \sqrt{4 + 6 (1 - 3 \epsilon) \epsilon} - 3 \epsilon \right),
\end{equation}

\begin{equation}
    C\equiv\frac{1}{6} \left(4  + \sqrt{4 + 6 (1 - 3 \epsilon) \epsilon} - 6 \epsilon \right),
\end{equation}

\begin{equation}
    D\equiv\frac{1}{6} \left(2  - \sqrt{4 + 6 (1 - 3 \epsilon) \epsilon} + 6 \epsilon \right),
\end{equation}

\begin{equation}
    E\equiv\frac{\sqrt{3}}{12}   \left(\sqrt{4 - 6 \epsilon} -  2 \sqrt{1 + 3 \epsilon}\right).
\end{equation}

In the NO case, when $M_{diag} = diag.(0, m_2e^{2i\alpha}, m_3)$, it can be shown that $\Delta S$ takes the following explicit form

\begin{align}
    \Delta S_{NO} = 1 - \bigg[\frac{1}{3m_2}\bigg]\Biggr(&\biggr[A^2 m_2^2+B^2m_3^2\sin^4\theta +2ABm_2m_3\sin^2\theta\cos2\alpha \biggr]^\frac{1}{2}
\nonumber \\
+&\biggr[C^2m_2^2 + D^2m^2_3\sin^4\theta + E^2m_3^2\sin^22\theta+2CDm_2m_3\sin^2\theta\cos2\alpha
\nonumber \\
& -2CEm_2m_3\sin2\theta\cos(2\alpha-\phi)-2DEm^2_3\sin^2\theta\sin2\theta\cos\phi\biggr]^\frac{1}{2}
\nonumber \\
+&\biggr[C^2m_2^2 + D^2m^2_3\sin^4\theta + E^2m_3^2\sin^22\theta+2CDm_2m_3\sin^2\theta\cos2\alpha 
\nonumber \\
&+2CEm_2m_3\sin2\theta\cos(2\alpha-\phi) +2DEm^2_3\sin^2\theta\sin2\theta\cos\phi\biggr]^\frac{1}{2}\Biggr). 
\nonumber \\
\end{align}

\noindent Similarly, in the IO case where $M_{diag} = diag.(m_1, m_2e^{2i\alpha}, 0)$, the measure of the degree to which the magic texture symmetry is broken is given by

\begin{align}
    \Delta S_{IO} = 1 - \bigg[\frac{1}{3m_2}\bigg]\Biggr(&\biggr[A^2 m_2^2+B^2m_3^2\cos^4\theta +2ABm_2m_3\cos^2\theta\cos2\alpha \biggr]^\frac{1}{2}
\nonumber \\
+&\biggr[C^2m_2^2 + D^2m^2_3\cos^4\theta + E^2m_3^2\sin^22\theta+2CDm_2m_3\cos^2\theta\cos2\alpha
\nonumber \\
& +2CEm_2m_3\sin2\theta\cos(2\alpha-\phi)+2DEm^2_3\cos^2\theta\sin2\theta\cos\phi\biggr]^\frac{1}{2}
\nonumber \\
+&\biggr[C^2m_2^2 + D^2m^2_3\cos^4\theta + E^2m_3^2\sin^22\theta+2CDm_2m_3\cos^2\theta\cos2\alpha 
\nonumber \\
&-2CEm_2m_3\sin2\theta\cos(2\alpha-\phi) -2DEm^2_3\cos^2\theta\sin2\theta\cos\phi\biggr]^\frac{1}{2}\Biggr). 
\nonumber \\
\end{align}

\begin{figure}[H]
    \centering
    \includegraphics[width=0.75\linewidth]{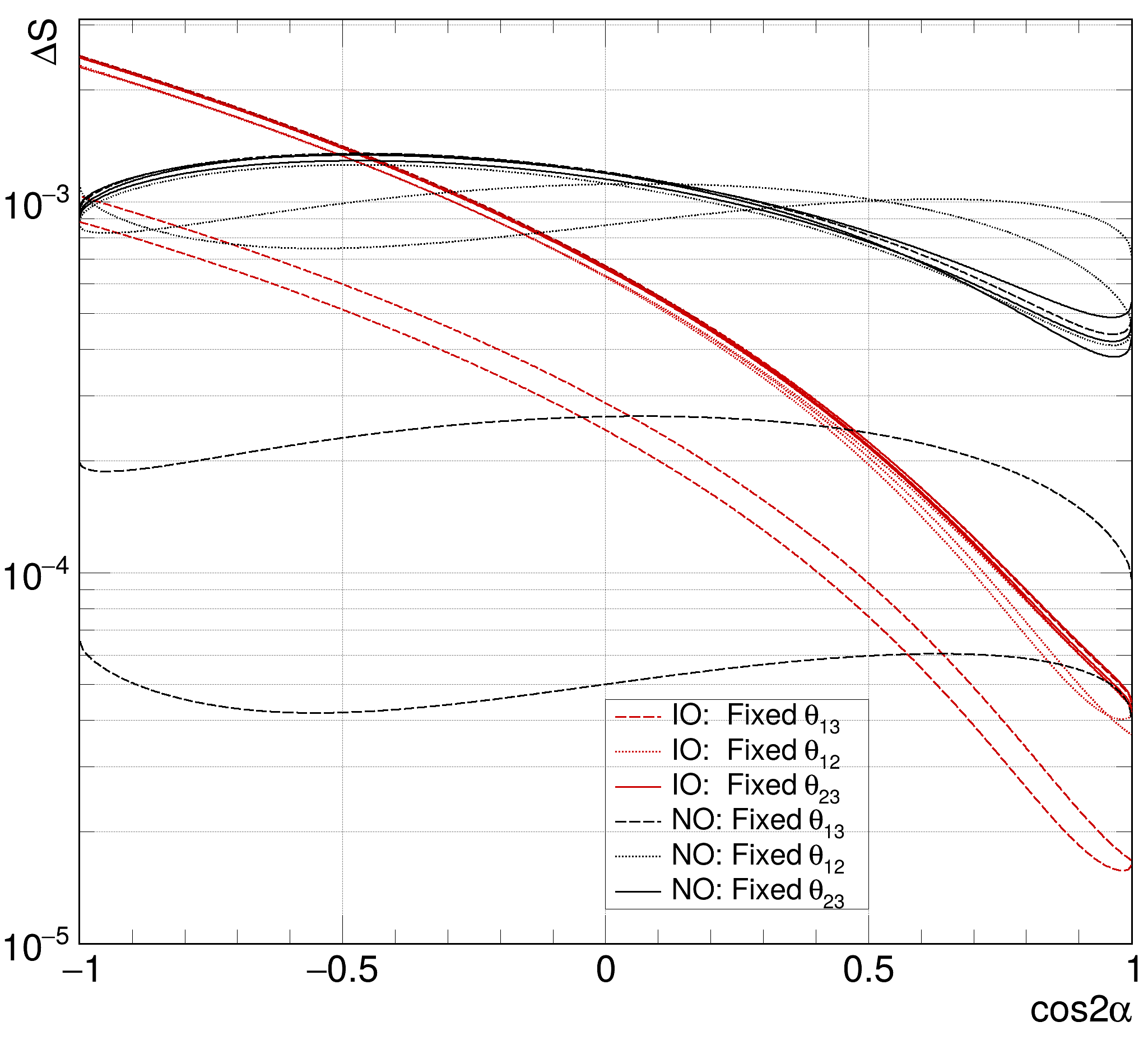}
    \caption{$\Delta S$ vs $\cos(2\alpha)$ in the NO(black) and IO(red) scenarios .}
    \label{fig:magic_texture_symmetry_violation}
\end{figure}

\noindent Using these equations we have explored the degree of violation of magic texture symmetry present for the same selection of $\theta/\phi/\epsilon$ values as was used in \ref{subsubsection:Majorana CP phases}. Done in this way we may get a sense of the region of parameter space covered by points within the NuFIT $3\sigma$ region in each of the fixed-angle scenarios. This can be seen for both the NO and IO cases in Figure \ref{fig:magic_texture_symmetry_violation}. 

Here the curves in the NO and IO case which have the smallest average value correspond to the smallest value of $\epsilon$, as expected. It may also be seen that for small values of $\alpha$ the magic texture symmetry is preserved best. If it is indeed more natural that the violation of this symmetry is minimized, and that in turn $\alpha$ should be small, then the expected value of $m_{\beta \beta}$ in the IO case should be well within the $90\%$ C.L. regions of both KamLAND-Zen and the planned next generation XLZD experiment. On the contrary in the NO case, even for small $\alpha$, $m_{\beta \beta}$ is expected to fall outside the $90\%$ C.L. region for GERDA, KamLANDm and XLZD, though in the case of XLZD only marginally. 

\section{Summary \label{section:summary}}

As the precision of measurements of mixing angles from oscillation experiments continues to increase, so to does the likelihood that the primary requirement of any neutrino mixing model will be that of precisely reproducing those angles. With this in mind, we have shown that a modified form of the previously favored TM$_2$ mixing matrix will accomplish this goal. 

Furthermore, we have demonstrated that this model is robust against future changes to the best-fit values of the mixing angles by showing the range of mixing angles predicted by tuning the parameter of the modified TM$_2$ model. Additionally, selecting sets of model parameters producing mixing angles in the NuFIT 1/2/3$\sigma$ regions, we have used the modified TM$_2$ model to make predictions for the Majorana neutrino mass for neutrinoless double beta decay. These predictions are found to be generally in agreement with other such calculations, and indicate that the next generation of neutrinoless double beta decay experiments should be able to exclude the full inverted ordering region while not yet probing that of the normal ordering region. Lastly, we have investigated the degree to which our modifications break the magic texture symmetry present in the unmodified TM$_2$ mixng matrix. We have made the assumption that it is natural to minimization the deviation from this symmetry and noted that this generally occurs for small values of the Majorana phase, $\alpha$. In turn we have shown that in the IO case, assuming such small values for $\alpha$, that $m_{\beta \beta}$ should be well within the $90\%$ C.L. region probed by both the current KamLAND-Zen and upcoming XLZD $0\nu\beta\beta$ experiments. On the contrary, for all ranges of $\alpha$, it is expected that the $m_{\beta\beta}$ is to remain outside the $90\%$ C.L. region for all curent and planned such experiments - although for small $\alpha$ only marginally outside of the range of XLZD.
Our modified TM$_2$ is a successful model in that it accomplishes our goal of precisely reproducing the current best-fit vales of the neutrino mixing angles while being robust to changes - but we acknowledge that further investigations must be made into a more fundamental way of deriving this matrix.


\section*{Acknowledgements\label{Section:Ackknowledgements}}
The authors would like to thank the anonymous referee for their constructive comments which have significantly improved the analysis and readability of this paper.


\vspace{1cm}

\end{document}